\documentclass[aps,prb,reprint,showpacs,superscriptaddress,amsmath,amssymb]{revtex4-1}

\usepackage{graphicx}
\usepackage{dcolumn} 
\usepackage{bm} 
\usepackage{hyperref} 
\usepackage{color}

\definecolor{red}{rgb}{1,0,0}

\definecolor{green}{rgb}{0,1,0}

\begin{document}

\title[]{Enhanced superconducting transition temperature in hyper-interlayer-expanded FeSe despite the suppressed electronic nematic order and spin fluctuations}

\author{Matev\v{z} Majcen Hrovat}
\affiliation{Jo\v{z}ef Stefan Institute, Jamova c. 39, SI-1000 Ljubljana, Slovenia}

\author{Peter Jegli\v{c}}
\affiliation{Jo\v{z}ef Stefan Institute, Jamova c. 39, SI-1000 Ljubljana, Slovenia}

\author{Martin Klanj\v{s}ek}
\affiliation{Jo\v{z}ef Stefan Institute, Jamova c. 39, SI-1000 Ljubljana, Slovenia}

\author{Takehiro Hatakeda}
\affiliation{Department of Applied Physics, Tohoku University, 6-6-05 Aoba, Aramaki, Aoba-ku, Sendai 980-8579,  Japan}

\author{Takashi Noji}
\affiliation{Department of Applied Physics, Tohoku University, 6-6-05 Aoba, Aramaki, Aoba-ku, Sendai 980-8579,  Japan}

\author{Yoichi Tanabe}
\affiliation{Department of  Physics, Tohoku University, 6-6-05 Aoba, Aramaki, Aoba-ku, Sendai 980-8579,  Japan}

\author{Takahiro Urata}
\affiliation{Department of  Physics, Tohoku University, 6-6-05 Aoba, Aramaki, Aoba-ku, Sendai 980-8579,  Japan}

\author{Khuong K. Huynh}
\affiliation{Department of  Physics, Tohoku University, 6-6-05 Aoba, Aramaki, Aoba-ku, Sendai 980-8579,  Japan}

\author{Yoji Koike}
\affiliation{Department of Applied Physics, Tohoku University, 6-6-05 Aoba, Aramaki, Aoba-ku, Sendai 980-8579,  Japan}

\author{Katsumi Tanigaki}
\affiliation{Department of  Physics, Tohoku University, 6-6-05 Aoba, Aramaki, Aoba-ku, Sendai 980-8579,  Japan}

\author{Denis Ar\v{c}on}
\email{denis.arcon@ijs.si}
\affiliation{Jo\v{z}ef Stefan Institute, Jamova c. 39, SI-1000 Ljubljana, Slovenia}
\affiliation{Faculty of Mathematics and Physics, University of Ljubljana, Jadranska c. 19, SI-1000 Ljubljana, Slovenia}

\begin{abstract}
The superconducting critical temperature, $T_{\rm c}$, of FeSe can be dramatically enhanced by intercalation of a molecular spacer layer. Here we report on a  $^{77}$Se, $^7$Li and $^1$H nuclear magnetic resonance (NMR) study of the powdered hyper-interlayer-expanded Li$_{x}($C$_2$H$_8$N$_2$)$_y$Fe$_{2-z}$Se$_2$  with a nearly optimal  $T_{\rm c}=45$~K. The absence of any shift in the $^7$Li and $^1$H NMR spectra indicates a complete decoupling of interlayer units from the conduction electrons in FeSe layers, whereas nearly temperature-independent $^7$Li and $^1$H spin-lattice relaxation rates are consistent with the non-negligible concentration of Fe impurities present in the insulating interlayer space. On the other hand, strong temperature dependence of  $^{77}$Se NMR shift and spin-lattice relaxation rate, $1/^{77}T_1$,  is attributed to the hole-like bands close to the Fermi energy. $1/^{77}T_1$ shows no additional anisotropy that would account for the onset of electronic nematic order down to   $T_{\rm c}$. Similarly, no enhancement in $1/^{77}T_1$ due to the spin fluctuations could be found in the normal state.  Yet, a characteristic power-law dependence $1/^{77}T_1\propto T^{4.5}$ still comply with the Cooper pairing mediated by spin fluctuations.  
\end{abstract}

\date{\today}


\maketitle

%
\section{Introduction}
%
%
The discovery of superconductivity in a layered iron oxypnictide\cite{Hosono} has triggered an intensive research activity to optimize the  unconventional superconducting properties of iron-based superconductors. Changing the composition of iron-based superconductors led to two distinct  families, iron pnictides and iron chalcogenides that share the same structural motif of electronically active layers composed of FeAs and FeQ (Q=Se,Te)  tetrahedra, respectively.    
A binary Fe$_{1+\delta}$Se adopts a particularly simple PbO-type structure,\cite{Hsu23092008} where the structural tetragonal-to-orthorhombic transition at $T_{\rm s}=91$~K (Ref. \onlinecite{baek_2015}) is, unlike as in 1111 or 122 FeAs compounds,\cite{dai_2012} not accompanied by the spin-density-wave (SDW) magnetic ordering. However, below   $T_{\rm s}$ the rotational $(C_4)$ symmetry is broken as the electronic nematic order\cite{dai_2012,Hu2012, Lu2014, fernandes_2014, chu_2012, watson_2015,  bohmer_2015} is established thus raising important questions regarding the absence of SDW, what triggers the electronic nematic order and what are its implications for the superconductivity in the iron-chalcogenide family. 

The two main candidates that may drive the electronic nematic order are electron spin and orbital degrees of freedom. Pronounced splitting of $^{77}$Se nuclear magnetic resonance (NMR) spectra in high quality FeSe single crystals below $T_{\rm s}$ has a characteristic order parameter temperature dependence and has been associated with the symmetry lowering due to the orbital ordering.\cite{baek_2015, bohmer_2015} The simultaneous absence of enhancement of the $^{77}$Se spin-lattice relaxation rate due to the spin fluctuations close to $T_{\rm s}$ implies that the orbital degrees of freedom drive the nematic order.\cite{bohmer_2015} High-resolution angle-resolved photoemission spectroscopy (ARPES) found Fermi surface deformations below $T_{\rm s}$ as a result of the splitting of bands associated with $d_{xz}$ and $d_{yz}$ character thus corroborating the orbital ordering scenario.\cite{watson_2015}  On the other hand, recent neutron scattering study revealed substantial stripe spin fluctuations that are coupled with orthorhombicity and enhanced close to $T_{\rm s}$, thus favoring spin fluctuations as the driving mechanism for the nematicity.\cite{wang_2015} This later possibility also seems to be more consistent with the theoretical studies that predict nematic quantum paramagnetic state\cite{kievelson_2015} with spin fluctuations at $\bf{q}=(\pi,Q)$ (where $Q=0$, $\pi/4$, $\pi/3$, $\pi/2$, ...).\cite{glasbrenner_2015} The high energy of spin fluctuations renders them unobservable by NMR thus explaining the absence of significant enhancement of spin-lattice relaxation rate close to $T_{\rm s}$.       

Fe$_{1+\delta}$Se is a superconductor with a critical temperature $T_{\rm c} \approx 8$~K at ambient
pressure.\cite{Hsu23092008, Serena2008, Cava2009, bohmer_2015, baek_2015}   With the application of hydrostatic pressure $T_{\rm c}$ dramatically increases reaching the maximum of $37$~K at $\sim 7$~GPa.\cite{Serena_2009} Strikingly,  single FeSe layers grown on SrTiO$_3$ show superconductivity at even higher temperatures, in some cases at critical temperatures that  exceed 100~K.\cite{tan_2013, ge_2014} 
Significant enhancement of $T_{\rm c}$ is also observed in  FeSe structures intercalated with alkali metal coordinated to molecular spacers (e.g., ammonia, pyridine, ethylenediamine or hexamethylenediamine),\cite{scheidt_2012, Koike, burrard_2013, zheng_2013, Noji20148, Hosono_2014, Clarke_2014, Sun_2015} where $T_{\rm c}$ first nearly linearly increases with  increasing interlayer spacing $d$  between 5 and 9~\AA ,   and then roughly saturates at $T_{\rm c}\approx 45$~K for $d>9$~\AA .\cite{Noji20148} 
The degree of Fe vacancies in the FeSe layer and the non-negligible amount of Fe intercalated between FeSe layers  were found to influence the superconducting properties of lithium iron selenide hydroxides Li$_{1-x}$Fe$_x$(OH)Fe$_{1-y}$Se.\cite{Sun_2015} Detailed {\em ab initio} calculations indeed support this picture by finding that the Li atoms donate electrons to FeSe layer thus tuning the critical temperature.\cite{abinitio_2015}  Interestingly, Fe ions present in the layers separating FeSe layers may ferromagnetically order at low temperature\cite{Klauss_2015, wu_nmr_2015} thus reminiscing  the magnetic ordering of inter-layer rare-earth moments in the 1111 family, e.g. in NdOFeAs.\cite{NOFA}

\begin{figure}[tbp]
\includegraphics[width=0.85\linewidth]{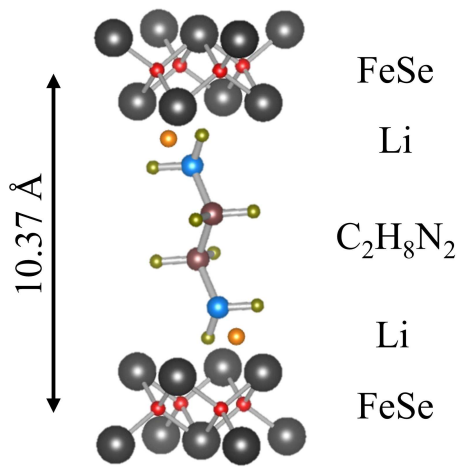}
\caption{\label{fig1} Schematic crystal structure of the intercalated FeSe-based  Li$_{x}($C$_2$H$_8$N$_2$)$_y$Fe$_{2-z}$Se$_2$ compound  with the hyper-expanded distance between neighboring Fe layers of $d=10.37$~\AA\,  and enhanced superconducting critical temperature $T_{\rm c}=45$~K. Here, red spheres represent Fe, gray spheres Se, orange spheres Li, blue spheres N, brown spheres C, and green spheres H. The position of impurity Fe atoms in the interlayer space is unknown.}
\end{figure}

How various factors, (i) disorder (ion vacancies, intercalated Fe or simply the structural disorder related to the co-intercalated molecular coordination), (ii)  dimensionality, (iii) spin fluctuations,  and (iv) possible nematicity vary across the intercalated-FeSe phase diagram is still unclear. Here we report  a systematic $^{77}$Se, $^7$Li and $^1$H NMR study of hyper-interlayer-expanded (lattice constant $c=20.74$~\AA\, yields  $d=10.37$~\AA) Li co-intercalated with ethylenediamine composition  Li$_{x}($C$_2$H$_8$N$_2$)$_y$Fe$_{2-z}$Se$_2$ (i-FeSe), Fig. \ref{fig1}, with a superconducting critical temperature $T_{\rm c}=45$~K.\cite{Koike} Probing the intra- and interlayer properties, we find that the studied i-FeSe with nearly optimal $T_{\rm c}$ should indeed be treated as a two-dimensional electronic system, which is not significantly perturbed by the considerable disorder present in the interlayer space. The absence of enhanced spin fluctuations and (within the resolution of powder NMR data) of nematic order provide important constrains for the superconducting state in the studied i-FeSe.   

%
%

%
\section{Experimental methods}
%
%
Samples were prepared according to standard procedures described in detail in Ref. \onlinecite{Koike}.  Powder x-ray diffraction confirmed that the samples were composed of Li- and ethylenediamine-co-intercalated Li$_{x}($C$_2$H$_8$N$_2$)$_y$Fe$_{2-z}$Se$_2$ with some minor unreacted FeSe impurity.  The low-field magnetization measurements disclosed two superconducting transitions, the first at $T_{\rm c}=45$~K belonging to i-FeSe  and the second due to non-intercalated FeSe impurity at $T_{\rm c}=8$~K.\cite{Koike}

$^{77}$Se ($I=1/2$)  NMR experiments were conducted in a magnetic field of 9.4~T. The reference Larmor frequency of  $\nu_{\rm L}(^{77}{\rm Se}) = 76.282$~MHz was determined from Me$_2$Se standard. A two-pulse Hahn-echo sequence $\pi /2 - \tau - \pi - {\rm echo}$ with a $\pi /2$ pulse length of $7\,  \mu {\rm s}$ and an  interpulse delay $\tau =50\, \mu {\rm s}$ was employed.  $^{7}$Li ($I=3/2$) and $^{1}$H ($I=1/2$) NMR measurements were performed in a magnetic field of 2.35~T at  Larmor frequencies $\nu_{\rm L}(^{7}{\rm Li}) = 38.85$~MHz (LiCl has been taken as a reference standard) and $\nu_{\rm L}(^{1}{\rm H}) = 99.95$~MHz. For the $^{7}$Li quadrupole nuclei a two-pulse solid-echo sequence $\pi /2 - \tau - \pi/2 - {\rm echo}$ with a $\pi /2$ pulse length of $4.4\,  \mu {\rm s}$ and an interpulse delay $\tau =30\, \mu {\rm s}$ has been used.  $^{1}$H ($I=1/2$) NMR frequency-swept spectra were recorded with the two-pulse Hahn-echo sequence ($\pi/2$ pulse length was $8$~$\mu{\rm s}$ and $\tau=20$~$\mu{\rm s}$).   The $^{77}$Se,  $^{7}$Li and  $^{1}$H spin lattice relaxation rates, $1/^{77}T_1$, $1/^{7}T_1$ and $1/^{1}T_1$,  were measured with inversion-recovery technique at the corresponding NMR lineshape peak positions.

%
\section{Results and discussion}
\begin{figure}[tbp]
\includegraphics[width=1.0\linewidth]{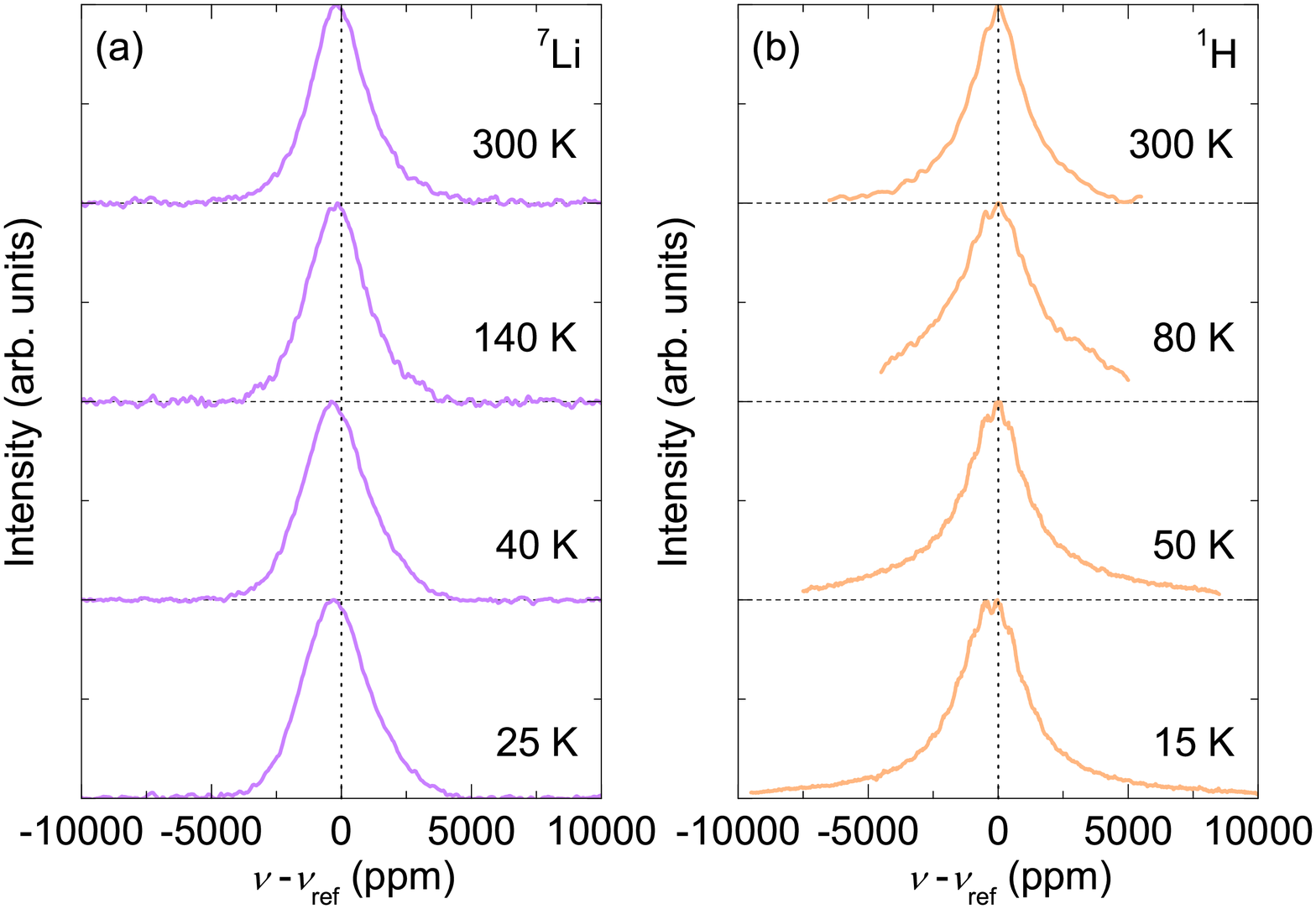}
\caption{\label{fig2} Temperature evolution of (a) $^7$Li and (b) $^1$H NMR spectra of i-FeSe powder. Please note the absence of shift and the similar line-broadening for both nuclei. Dotted vertical lines indicate the corresponding Larmor frequencies.  }
\end{figure}

%
%
The room-temperature $^1$H and $^7$Li NMR spectra  are  featureless, symmetric and centered close to their  Larmor frequencies (Fig. \ref{fig2}). The absence of a measurable shift in the $^7$Li NMR spectrum implies an almost fully ionized Li$^+$ species and a complete charge transfer to the FeSe layer. The linewidth of the $^1$H  NMR spectrum amounts to 2083(16)~ppm at 300 K, which is an expected line-broadening caused by the proton-proton dipolar interactions of co-intercalated ethylenediamine molecules. We note, that the  $^7$Li NMR spectrum  at 300 K exhibits a nearly identical linewidth of 2417(19)~ppm implying a similar size of the local magnetic fields at the Li site. Since Li atoms are, likewise to ethylenediamine molecules, intercalated between FeSe layers, we conclude that the proton dipolar fields also broaden $^7$Li NMR spectra. On cooling, both $^1$H and $^7$Li NMR spectra show only a very moderate broadening and do not shift away from their respective Larmor frequencies. For comparison, small but non-zero $^{23}$Na and $^7$Li NMR shifts due to the weak transferred hyperfine coupling to FeAs layer were found in NaFeAs and LiFeAs.\cite{NaFeAs, LiFeAs} Moreover, the $^1$H and $^7$Li NMR spectra  in i-FeSe are almost insensitive to the  superconducting transition at $T_{\rm c}=45$~K. These observations unambiguously prove that the Li atoms and ethylenediamine molecules feel no hyperfine field and are thus completely decoupled from the conducting electrons in FeSe layers. Their role is thus to provide charges to FeSe layer and to separate these layers, thus establishing  i-FeSe compound as a perfect two-dimensional conductor. 

\begin{figure}[tbp]
\includegraphics[width=1.0\linewidth]{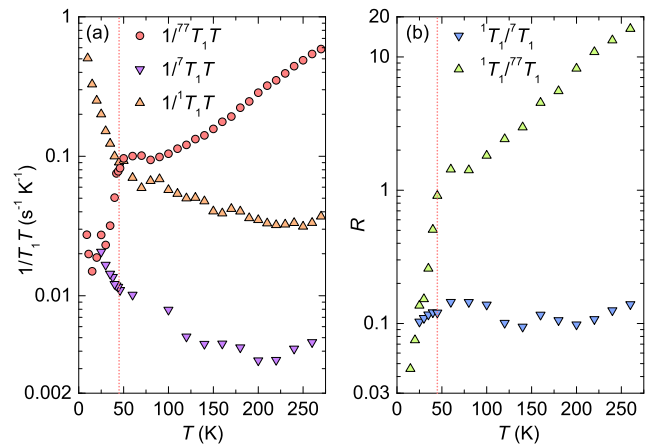}
\caption{\label{fig3} (a) Temperature dependences of  $^1$H (orange up triangles), $^7$Li (violet down triangles) and $^{77}$Se (red circles)  spin-lattice relaxation rates divided by temperature, $1/^{n}T_1T$ ($n=1, 7, 77$). (b) Temperature dependence of the ratio, $R$, of $^1$H to $^7$Li ,  $^1T_1/^7T_1$ (blue down triangles),  and of $^1$H to $^{77}$Se,    $^1T_1/^{77}T_1$ (green up triangles),  spin-lattice relaxation times. Dotted vertical lines indicate the superconducting critical temperature $T_{\rm c}=45$~K of i-FeSe sample.  }
\end{figure}

$^1$H and $^7$Li spin-lattice relaxation rates divided by temperature, $1/^{n}T_1T$ (here $n=1,7$ stands for $^1$H and $^7$Li, respectively), monotonically increase with decreasing temperature [Fig. \ref{fig3}(a)]. The ratio of $^1$H to $^7$Li spin-lattice relaxation times,  $^1T_1/^7T_1$, is nearly temperature independent   [Fig. \ref{fig3}(b)] thus proving that both nuclei experience the same spectrum of fluctuating local magnetic fields and corroborating  the conclusions derived from the $^1$H and $^7$Li  NMR spectra (Fig. \ref{fig2}). Evidently, there is no contribution from the relaxation governed by the hyperfine coupling to the conducting FeSe electrons. This explains why $1/^{n}T_1$ are not sensitive to the onset of superconductivity, which would otherwise   lead to a suppression of spin-lattice relaxation rates  below $T_{\rm c}$. 

Establishing the absence of conducting electron hyperfine fields in the interlayer space, one can expect that the spin-lattice relaxation in this  layer will be governed by other mechanisms usually encountered in insulators. In the case when spin-lattice relaxation is determined by the ethylenediamine molecular motions, the Bloembergen-Purcell-Pound (BPP)-type relaxation mechanism applies and $1/^{1}T_1$ should display strong thermally activated (Arrhenius-type) temperature dependence with a maximum at the temperature where the correlation time $\tau$ for the molecular motion matches the inverse Larmor frequency, i.e. when $\omega_{\rm L}\tau = 1$.\cite{CPS_1996, abragam} However, this is clearly not supported by nearly temperature independent $1/^{1}T_1$,  i.e. $1/^{1}T_1T$ roughly scales as $1/T$ [Fig. \ref{fig3}(a)] as between 300 K and 10~K  $1/^{1}T_1$ increases only for a factor of $\sim 2$.  Therefore, we conclude that the $^1$H (and also $^7$Li) spin-lattice relaxation process is short-cut by another        weakly temperature dependent relaxation mechanism. A plausible possibility that provides such a nearly temperature-independent spin-lattice relaxation   is a nuclear-spin diffusion toward the diluted localized magnetic moments.\cite{diff2} In two-dimensional diluted paramagnets $1/^{1}T_1\propto N_{\rm p}D^{3/4}$, where $N_{\rm p}$ is a concentration of paramagnetic impurities and $D$ is the nuclear spin diffusion constant.\cite{diff1}   
In i-FeSe, such diluted paramagnetic impurities could be associated with the non-negligible concentration of Fe impurities present in the insulating interlayer space. 
However, judging from the absence of additional broadening of $^1$H and $^7$Li NMR spectra and the absence of  $1/^{1}T_1$ enhancement these Fe impurities show no tendency towards magnetic ordering down to 10~K.

\begin{figure}[tbp]
\includegraphics[width=1.0\linewidth]{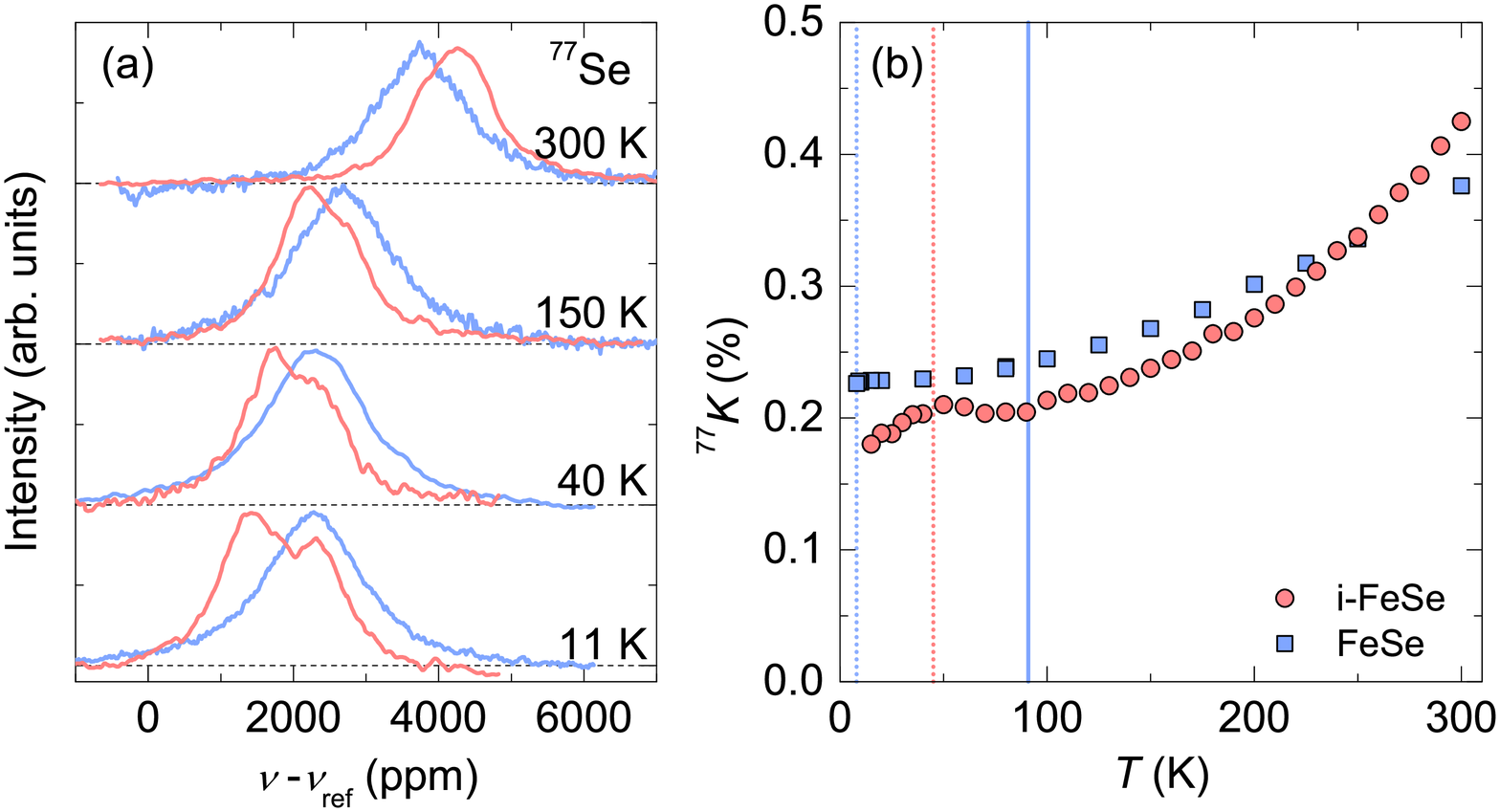}
\caption{\label{fig4} (a) Temperature evolution of the $^{77}$Se NMR spectra measured in i-FeSe (red line) and in FeSe (blue line) powders. Note the presence of a second weak peak attributed to the presence of non-intercalated FeSe impurity regions in the spectra of   i-FeSe. 
(b) Temperature dependences of the $^{77}$Se NMR  shifts for  i-FeSe (red circles) and FeSe (blue squares).
Dotted vertical lines indicate the superconducting critical temperature $T_{\rm c}=45$~K for i-FeSe (red) and $T_{\rm c}=8$~K for FeSe (blue). Solid vertical blue line marks a FeSe structural phase transition at $T_{\rm s}=91$~K.  }
\end{figure}

Since Li and ethylenediamine species are electronically completely isolated from FeSe layers, we now turn to the  $^{77}$Se NMR to directly probe  the electronic properties of the FeSe layer. 
Comparison of  the $^{77}$Se NMR spectra of powdered FeSe and i-FeSe samples is shown in Fig. \ref{fig4}(a). The linewidth of the i-FeSe spectrum is 90 kHz (which corresponds to $\sim 1200$~ppm) at 300~K and is comparable to  that of the FeSe powder.  The  broadening due to the $^1$H-$^{77}$Se dipolar interactions is estimated to be negligible compared to the hyperfine broadening between the $^{77}$Se nuclear moments and the conducting electrons in the FeSe layers.  In addition to the main  $^{77}$Se NMR line of i-FeSe, a second minor peak becomes more pronounced below $\sim 150$~K.  A direct comparison with the spectra of FeSe reveals that this weaker resonance   in fact belongs to the small amount of non-intercalated FeSe regions present in our sample.   

The shift of the $^{77}$Se  NMR spectra, $^{77}K$, in i-FeSe corresponds to 4250~ppm at room temperature and is significantly larger compared to FeSe where it amounts to 3760~ppm. On cooling,  $^{77}K(T)$ shows a very strong temperature dependence [Fig. \ref{fig4}(b)], roughly following the empirical  $^{77}K(T)=K_0+k T^2$ dependence with the fitting constants $K_0=1834$~ppm and $k =0.025$~ppm/K$^2$. On the other hand, FeSe shows weaker temperature dependence of  $^{77}K$ and, if analyzed with the same empirical model,  a significantly smaller $k = 0.017$~ppm/K$^2$ (and  larger $K_0=2279$~ppm)  is obtained. In general, the shift is composed of the temperature-independent orbital contribution, $K_{\rm orb}$, and the Knight shift, $K_{\rm s}$, respectively. The Knight shift $K_{\rm s}$ is related to the density of states $g(\epsilon )$  in the vicinity of the Fermi energy $\epsilon_{\rm F}$, i.e.,
\begin{equation}
K_{\rm s}(T)\propto \int g(\epsilon )\left( -{\partial f\over \partial \epsilon}\right)   d\epsilon ,
\end{equation}
where $f(\epsilon)=1/(1+\exp[(\epsilon-\epsilon_{\rm F})/k_{\rm B}T])$ is the Fermi-Dirac distribution function. When $\epsilon_{\rm F}$ is positioned somewhere in the middle of the conduction band, where $g(\epsilon)$ does not change significantly over the energy range of $k_{\rm B}T$, the above expression  predicts temperature independent $K_{\rm s}$, which is proportional to the density of states at the Fermi level $g(\epsilon_{\rm F})$. However,  $K_{\rm s}$ can become  temperature dependent, if $g(\epsilon )$ changes substantially in the energy interval $|\epsilon -\epsilon_{\rm F}|\sim k_{\rm B}T$. In FeSe, this condition seems to be fulfilled for the hole-like pockets $\alpha$, $\beta$ and $\gamma$ with band-edges very close to $\epsilon_{\rm F}$ according to the recent ARPES study.\cite{watson_2015} Although no comparable ARPES study is at the moment available for our i-FeSe sample, we can qualitatively argue that the much steeper and stronger temperature dependence of $K_{\rm s}$ in i-FeSe suggests, that at least one of these bands is pushed even closer to $\epsilon_{\rm F}$ upon intercalation.  

\begin{figure}[tbp]
\includegraphics[width=1.0\linewidth]{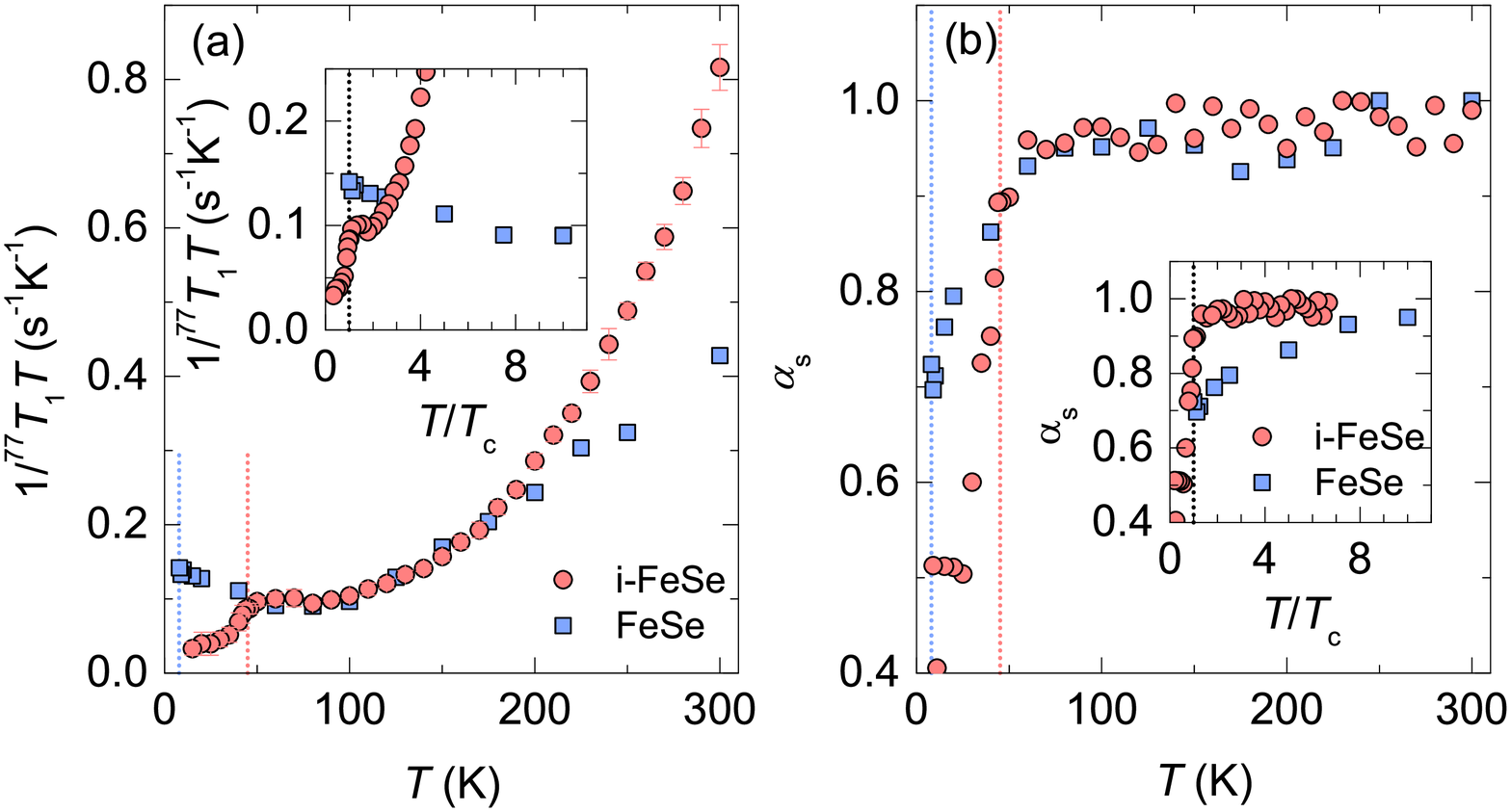}
\caption{\label{fig42} Temperature dependences  of the  $^{77}$Se spin-lattice relaxation rates, $1/^{77}T_1$, divided by temperature (a) and of the stretching exponent $\alpha_{\rm s}$ (b) for  i-FeSe (red circles) and FeSe (blue squares). Insets to (a) and (b) show the temperature dependences of $1/^{77}T_1T$ and $\alpha_{\rm s}$ in the reduced $T/T_{\rm c}$ scale.
Dotted vertical lines indicate the superconducting critical temperature $T_{\rm c}=45$~K for i-FeSe (red) and $T_{\rm c}=8$~K for FeSe (blue). Dotted vertical black lines in the insets  mark the onset of superconductivity where $T/T_{\rm c}=1$.  }
\end{figure}

A very similar conclusion is derived also from the  $^{77}$Se spin-lattice relaxation rates [Fig. \ref{fig42}(a)]. The Korringa-type relaxation 
\begin{equation}
1/T_1T\propto \int g^2(\epsilon )\left( -{\partial f\over \partial \epsilon}\right)   d\epsilon ,
\end{equation}
predicts temperature independent $1/T_1T$ rates only for the cases when band edges are far away from  $\epsilon_{\rm F}$. Therefore, a very steep increase of $1/^{77}T_1T$ with increasing temperature for $T>100$~K implies that the relaxation rate is enhanced because at least one of the the hole-like pockets $\alpha$, $\beta$ and $\gamma$ has features very close to $\epsilon_{\rm F}$. Again, the temperature dependence of $1/^{77}T_1T$  in i-FeSe is more pronounced than in FeSe thus corroborating the conclusions derived from the  discussion of the Knight shift data.  

Slight shifting of the hole-like pockets $\alpha$, $\beta$ and $\gamma$ in i-FeSe compared to FeSe samples may  also be responsible for an important difference between the temperature dependencies of $1/^{77}T_1T$ in the two samples below $T\approx 100$~K.  Namely, in FeSe $1/^{77}T_1T$ starts to increase with decreasing temperature in the electronic nematic phase below the structural phase transition $T_{\rm s}=91$~K [Fig. \ref{fig42}(a)], in agreement with the literature data.\cite{baek_2015, bohmer_2015, imai_2009} Such an enhancement of the relaxation rate is a strong indication of the enhancement in the spectral density function of spin fluctuations at the nuclear Larmor frequency. The increase in $1/^{77}T_1T$ is accompanied by the larger distribution of $1/^{77}T_1$ values, because of the larger anisotropy in  $1/^{77}T_1$ in the electronic nematic phase.\cite{baek_2015, bohmer_2015} In the present experiments on  powdered samples this is evident from the stretched-exponential form of the $^{77}$Se nuclear magnetization recovery data. The stretching exponent $\alpha_{\rm s}$ decreases from 0.96(8) for $T_{\rm s}<T<300$~K to 0.73(5) at $T_{\rm c}=8$~K [Fig. \ref{fig42}(b)].  Returning back to the i-FeSe sample, we notice that both, the enhancement in $1/^{77}T_1T$ as well as the decrease in $\alpha_{\rm s}$ are completely absent down to the superconducting critical temperature $T_{\rm c}=45$~K. 
Since the superconducting critical temperature defines the appropriate energy scale, we next plot both parameters in the reduced $T/T_{\rm c}$ scale (insets to Fig. \ref{fig42}). Whereas in FeSe the enhancement in $1/^{77}T_1T$ and the suppression of   $\alpha_{\rm s}$ can be tracked up to $T/T_{\rm c}\approx 10$, the behavior in i-FeSe is markedly different.
 More specifically, in i-FeSe   $1/^{77}T_1T$ and $\alpha_{\rm s}$ remain nearly temperature-independent at 0.097(8)~s$^{-1}$K$^{-1}$ and 0.95(5) between  $T/T_{\rm c}=1$ and $T/T_{\rm c}\approx 2$, respectively. This   demonstrates  the suppression of spin fluctuations probed at the NMR frequency and the absence of  anisotropy in the electronic response, the later speaking for the suppression of electronic nematic order, in the normal state of i-FeSe.

\begin{figure}
\includegraphics[width=1.0\linewidth]{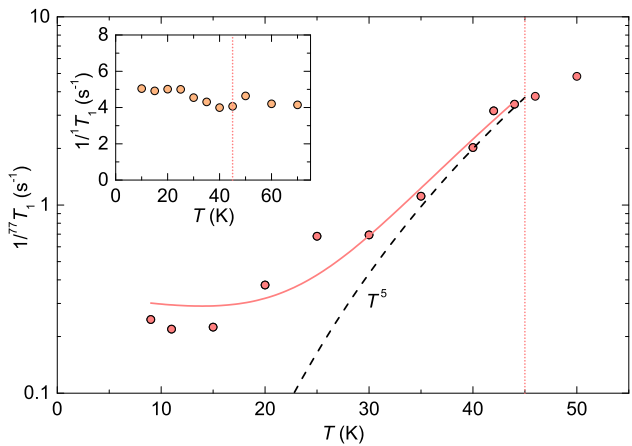}
\caption{\label{fig5} (a) Temperature dependence of $^{77}$Se spin-lattice relaxation rate, $1/^{77}T_1$, below the i-FeSe superconducting critical temperature $T_{\rm c}=45$~K. The dashed line shows the expected temperature dependence for a power-law   $1/^{77}T_1\propto T^{5}$ compatible with the $s^\pm$ superconductivity. Solid red line is a fit to $1/^{77}T_1=A+B(T/T_{\rm c})^n$ with $A=0.24$~s$^{-1}$, $B=3.39$~s$^{-1}$ and $n=4.5$. Inset: $^1$H spin-lattice relaxation rate, $1/^{1}T_1$, is nearly temperature-independent below $T_{\rm c}$. Thin dotted vertical lines mark $T_{\rm c}$.}
\end{figure}

Comparative NMR study of i-FeSe and FeSe in the normal state discloses some important electronic differences between the two compounds: (i) at least some of the hole-like pockets in i-FeSe shift closer to $\epsilon_{\rm F}$, (ii) there is a suppression of spin fluctuations probed at the nuclear Larmor frequency and (iii) (within the resolution of present powder experiments) there is also a suppression of nematicity in i-FeSe. These differences are expected to be reflected also in the superconducting state. 
For instance, if spin fluctuations mediate the Cooper pairing, the system is predicted to develop  an unconventional $s^{\pm}$ superconductivity.\cite{splusminus} On the other hand, when orbital fluctuations are in play, then  $s^{++}$ state is predicted.\cite{splusplus} 
Moreover, since it was suggested that the nematicity competes with the superconductivity in FeSe,\cite{baek_2015} our finding of  suppressed electronic nematicity  may at least qualitatively account for the high $T_{\rm c}$  in i-FeSe.  
In Fig. \ref{fig5} we show the low-temperature dependence of $1/^{77}T_1$.  A sharp suppression of $1/^{77}T_1$ below $T_{\rm c}$ confirms the opening of the superconducting gap and reveals the absence of a characteristic spin-lattice relaxation rate enhancement due to the coherence peak just below $T_{\rm c}$.  We stress that for the $s^{++}$ superconducting state the coherence peak would be expected. However, as it has been  observed  in isotropic $s-$wave strongly correlated systems\cite{choi, Tone} the damping effects arising from the scattering of the electron with other electrons  may suppress the coherence peak. Although electron correlations have been discussed in iron-chalcogenides,\cite{Medici, AnaTamai, FST} it is still unlikely that they are strong enough to account for the experimental observation of the completely suppressed  coherence peak. Therefore, we consider the orbital fluctuations $s^{++}$ scenario in i-FeSe less probable.

Below $T_{\rm c}$,  $1/^{77}T_1$ adopts a power-law temperature dependence,  $1/^{77}T_1\propto T^n$. Such a power-law dependence can be found for most iron-pnictides\cite{NMRpow2, NMRpow1, NMRpow3, NMRpow4, Yashima} and iron-chalcogenides\cite{FeSeNMR} with $n$ varying from $3-5$. Various models based on   $s^{\pm}$-wave symmetry with the two Fermi surfaces dominated by an isotropic full and an anisotropic full gap can account for such dependence.\cite{Yashima} Indeed, fitting   $1/^{77}T_1$  of i-FeSe below $T_{\rm c}$ to $1/^{77}T_1\propto T^n$ with $n\approx 5$ provides a satisfactory fit of the data (Fig. \ref{fig5}). 
However, at the lowest temperatures below $\sim 15$~K, when the main relaxation channel via thermally excited quasiparticles becomes very weak, $1/^{77}T_1$ suddenly tends to saturate. This is reminiscent of nearly temperature independent $^1$H (and also $^7$Li) relaxation rates (inset to Fig. \ref{fig5}) thus implying that very weak  fluctuating  fields originating from the  interlayer impurity Fe magnetic moments provide an additional relaxation channel for the $^{77}$Se nuclei. Therefore, we fit $1/^{77}T_1$  to a sum of two contributions, $1/^{77}T_1=A+B(T/T_{\rm c})^n$, for all $T<T_{\rm c}$. Here, $A$ and $B$ are the fitting constants related to  the magnitude of the fluctuating  insulating interlayer moments and to the hyperfine fields from the electrons in the FeSe layer, respectively. Finally, the extracted  $n=4.5$ is in qualitative agreement with the $s^{\pm}$ scenario.  

We remark that the presence of additional temperature-independent  relaxation channel in i-FeSe  introduces some uncertainty to the extracted fitting parameters in the superconducting state. 
Nevertheless, most of the NMR data still seems to be compatible with the Cooper pairing  mediated by spin fluctuations. This may be  in apparent contradiction    with the observation of suppressed spin-fluctuations in the normal state [Fig. \ref{fig42}(a)]. However, if the theoretical suggestion that spin fluctuations are very high in energy is correct,\cite{kievelson_2015} then their influence on the nuclear spin relaxation would be negligible and could still account for the very high $T_{\rm c}$ in i-FeSe. Our finding that the hole-like pockets shift closer to $\epsilon_{\rm F}$  in i-FeSe may then hold important clues about the tuning of spin fluctuations and optimizing $T_{\rm c}$ in this family of materials.

In conclusion,  Li$_{x}($C$_2$H$_8$N$_2$)$_y$Fe$_{2-z}$Se$_2$  with a superconducting critical temperature $T_{\rm c}=45$~K has been studied with $^{77}$Se, $^7$Li and $^1$H NMR. Electronically active FeSe layers are found to be completely decoupled from each other by insulating layers comprising  ethylenediamine, Li and, importantly, also intercalated (impurity) Fe atoms. In comparison to the parent FeSe, i-FeSe shows completely suppressed electronic nematicity and the absence of spin fluctuations  probed at the NMR Larmor frequency. However, the Cooper pairing  mediated by high-energy spin fluctuations still provides the  best explanation for the absence of the coherence peak and the power-law dependence of the spin-lattice relaxation rates below $T_{\rm c}$. The family of intercalated FeSe compounds thus emerges as an intriguing case where the intertwining of lattice, charge and  spin degrees of freedom establishes a highly intricate superconducting state with surprisingly high $T_{\rm c}$.   

%
%

%

%
\begin{acknowledgments}
D.A. acknowledges the financial support from the European Union FP7-NMP-2011-EU-Japan project LEMSUPER under Contract No. NMP3-SL-2011-283214 and from the Slovenian Research Agency project under Contract No. BI-JP/12-14-003.

\end{acknowledgments}
%
%

%
%
%
\bibliography{iFeSe}
\bibliographystyle{apsrev}

\end{document}